# A Complete Anytime Algorithm for Treewidth

**Vibhav Gogate and Rina Dechter**
School of Information and Computer Science,
University of California, Irvine, CA 92967
{vgogate,dechter}@ics.uci.edu

## Abstract

In this paper, we present a Branch and Bound algorithm called *QuickBB* for computing the treewidth of an undirected graph. This algorithm performs a search in the space of perfect elimination ordering of vertices of the graph. The algorithm uses novel pruning and propagation techniques which are derived from the theory of graph minors and graph isomorphism. We present a new algorithm called minor-min-width for computing a lower bound on treewidth that is used within the branch and bound algorithm and which improves over earlier available lower bounds. Empirical evaluation of QuickBB on randomly generated graphs and benchmarks in Graph Coloring and Bayesian Networks shows that it is consistently better than complete algorithms like Quick-Tree [Shoikhet and Geiger, 1997] in terms of cpu time. QuickBB also has good anytime performance, being able to generate a better upper bound on treewidth of some graphs whose optimal treewidth could not be computed up to now.

## 1 Introduction

Given an undirected graph $G$ and an integer $k$ the problem whether the treewidth of $G$ is at most $k$ is known to be NP-complete [Arnborg et al., 1987]. In this paper, we develop a complete anytime algorithm based on branch and bound search to solve the optimization version of this problem. In other words, we are interested in the smallest such $k$. The problem can also be rephrased as finding a *triangulation or tree-decomposition $T$* of $G$ such that the treewidth of $T$ is same as $G$.

A solution to this problem is important because many algorithms that solve NP-hard problems in Artificial Intelligence, Operations Research, Circuit Design etc. are exponential only in the treewidth. Examples of such algorithm are Bucket elimination [Dechter, 1999] and junction-tree elimination [Lauritzen and Spiegelhalter, 1988] for Bayesian Networks and Constraint Networks. These algorithms operate in two steps: (1) constructing a good tree-decomposition and (2) solving the problem on this tree-decomposition, with the second step generally exponential in the treewidth of the tree-decomposition computed in step 1. In this paper, we present a complete anytime algorithm called QuickBB for computing a tree-decomposition of a graph that can feed into any algorithm like Bucket elimination [Dechter, 1999] that requires a tree-decomposition.

The majority of recent literature on the treewidth problem is devoted to finding constant factor approximation algorithms [Amir, 2001, Becker and Geiger, 2001]. Other heuristic algorithms include popular triangulation heuristics like min-degree, max-cardinality search and min-fill. These heuristics have no performance guarantees and the approximation computed by them can be exponentially bad. We discuss these heuristics in the Section 3.

It is well known that the speed of a branch and bound algorithm depends upon the quality of the lower bound used. In this paper, we have developed a new lower bound called minor-min-width that is shown to produce a better lower bound than max-cardinality search developed recently by Brian Lucena [Lucena, 2003]. We implemented a branch and bound algorithm using minor-min-width as lower bound. However this implementation did not terminate on some graphs having 20 vertices in 2 days of cpu time. Some analysis revealed that this naive algorithm can be improved with some clever modifications along two directions: (1) Reducing the branching factor at each state, (2) Using pruning and propagation rules to prune regions of the search space that can not be a part of a triangulation having a lower treewidth than the current best.

We reduce the branching factor by using Dirac's theorem which characterized triangulated graphs and graph minor theory. Also, we derive various pruning and propagation rules from the theory of graph minors and graph isomorphism. We call the resulting algorithm QuickBB.

Prior to our work, the best existing complete algorithm for finding optimal triangulation is due to Shoikhet and Geiger(called QuickTree [Shoikhet and Geiger, 1997]). Our empirical evaluation on randomly generated graphs shows that QuickBB is superior to QuickTree. On some randomly generated graphs with 100 vertices and treewidth bounded by 10, QuickBB appears to be 50 times faster than QuickTree. Being a branch and bound algorithm, QuickBB also has good anytime properties. We were able to obtain better upper bounds on treewidth of some graphs in Dimacs Graph coloring benchmarks and Bayesian Network repository even when the algorithm did not terminate.

## 2  Definitions and Preliminaries

Note that all the lemmas presented in this section can be found in a book on treewidth by Ton Kloks [Kloks, 1994]. All the graphs used in this paper are finite, undirected, connected and simple. We denote an undirected graph by $G(V_G, E_G)$ where $V_G$ is the set of vertices of the graph and $E_G$ is the set of edges of the graph. The *neighborhood* of a vertex $v$ denoted $N_G(v)$ is the set of vertices that are neighbors of $v$. A chord of a cycle $C$ is an edge not in $C$ whose endpoints lie in $C$. A *chordless cycle* in $G$ is a cycle of length at least 4 in $G$ that has no chord.

**Definition 2.1.** *A graph is* **chordal** *or triangulated if it has no chordless cycle.*

A graph $G' = (V', E')$ is called a *subgraph* of $G$ if $V' \subseteq V$ and $E' \subseteq E$. If $W \subseteq V$ is a subset of vertices then $G[W]$ denotes the subgraph induced by $W$. A clique is a graph $G$ that is completely connected. A *triangulation* of a graph $G$ is a graph $H$ such that $G$ is a subgraph of $H$, $H$ has the same set of vertices as $G$ and $H$ is chordal.

**Definition 2.2.** *A* **tree decomposition** *of a graph $G(V, E)$ is a pair $(T, \chi)$ where $T = (I, F)$ is a tree and $\chi = \{\chi_i \mid i \in I\}$ is a family of subsets of $V$ such that: (1)$\bigcup_{i \in I} \chi_i = V$, (2)For each edge $e = \{u, v\} \in E$ there exists an $i \in I$ such that both $u$ and $v$ belong to $\chi_i$ and (3) For all $v \in V$, there is a set of nodes $\{i \in I | v \in \chi_i\}$ forms a connected subtree of $T$.*

The width of a tree-decomposition is given by: $max_{i \in I} (|\chi_i| - 1)$. The treewidth of a graph $G$ denoted by $tw(G)$ equals the minimum width over all possible tree decompositions of $G$. The treewidth of $G$ is the minimum $k \geq 0$ such that $G$ is a subgraph of a triangulated graph $H$ having a maximal clique of size $k + 1$.

**Definition 2.3.** *A vertex $v$ of $G$ is* **simplicial** *if its neighborhood induces a clique. A vertex $v$ is* **almost simplicial** *if all but one of its neighbors induce a clique. An ordering of the vertices $\pi = [v_1, v_2, \ldots, v_n]$ is called a* **perfect elimination ordering** *if for every $1 \leq i \leq n$, $v_i$ is a simplicial vertex in $G[X]$ where $X = \{v_i, \ldots, v_n\}$.*

**Definition 2.4.** *A clique of size $k + 1$ is a* **k-tree** *of size $k + 1$. A k-tree of size $n + 1$ can be constructed from a k-tree of size $n$ by taking the new vertex and making it adjacent to any clique of size $k$ in the k-tree. A subgraph of a k-tree is called a* **partial k-tree**.

**Lemma 2.5.** *The treewidth of a k-tree is $k$. The treewidth of a partial k-tree is at most $k$.*

A characterization of the triangulated graphs is given by the following lemma.

**Lemma 2.6.** *A graph is triangulated iff there exists a perfect elimination ordering for it. Also, if a graph is triangulated then any simplicial vertex can start a perfect elimination ordering for it.*

Another important property is due to Dirac.

**Lemma 2.7.** *Any non-clique triangulated graph has at least two non-adjacent simplicial vertices.*

The relation between the treewidth of a triangulated graph and perfect elimination ordering is captured by the following lemma.

**Lemma 2.8.** *Given any perfect elimination ordering $f = [v_1, \ldots, v_n]$ of a chordal graph $T$, the treewidth of $T$ is given by $max(|N(v_i)| \mid v \in V, N(v) \cap \{v_i, \ldots, v_n\})$*

A graph $M$ is a *minor* of a graph $G$ if graph $M$ can be obtained from a subgraph of $G$ by edge contraction. Edge contraction of an edge $e = \{u, v\}$ is the operation of replacing both $u$ and $v$ by a single vertex $w$ such that the neighbors of $u$ and $v$ are neighbors of $w$ except $u$ and $v$ themselves. A relation between graph, its minor and the treewidth of $G$ is given by the following lemma.

**Lemma 2.9.** *Let $G$ be a graph and $M$ be a minor of $G$. Then $tw(G) \geq tw(M)$.*

We say that we *eliminate* a vertex $v$ when we make $v$ simplicial and remove it from the graph to obtain a new graph $G'$. This operation will be denoted by $elim(G, v)$ and it returns a graph $G' = (V \setminus v, E \cup E')$, where $E' = \{(v_1, v_2) | v_1, v_2 \in N(v)\}$. Given a linear ordering $f = [v_1, \ldots, v_n]$ of an undirected graph $G$, a triangulation $T$ of the $G$ along the ordering can be obtained as follows. Make each vertex $v_i$ simplicial by connecting all its neighbors in the graph induced by $G[V]$, *where* $V = \{v_i, \ldots, v_n\}$. It is easy to see that the ordering $f$ is a perfect elimination ordering of $T$. The treewidth of a graph $G$ along an ordering $f$ is the treewidth of the triangulation $T$ obtained by triangulating the graph along the ordering $f$.

## 3 Popular Heuristics

In this section, we describe three popular heuristics for computing an upper bound on treewidth. These heuristics run in time polynomial in the size of the graph and therefore form important candidates that could be used to place good upper bounds on the branch and bound scheme. All the heuristics described below are used to construct a perfect elimination ordering of the graph $G$.

**The min-fill heuristic:** Order the vertices from 1 to $n$ as follows. First select a vertex $v$ which adds the least number of edges when eliminated from the graph and place it at position 1. Eliminate $v$ from the graph by making it simplicial. Now select any vertex that adds the least number of edges when eliminated and place it at the next position in the ordering. Repeat the process breaking ties arbitrarily.

**The min-width heuristic:** Order the vertices from 1 to $n$ as follows. First, select a vertex $v$ which has minimum degree and place at position 1. Then remove this vertex $v$ from the graph and select any vertex with minimum degree and place it at the next position in the ordering. Repeat this process breaking ties randomly.

**The max-cardinality heuristic:** We are ordering $n$ vertices of a graph $G$ from $n$ to 1. Label a random vertex as 1 and place it at position $n$. Choose as the next to label an unlabeled vertex $v$ with a maximum number of previously labeled neighbors breaking ties arbitrarily. Place this vertex at the next position in the ordering and label the vertices consecutively. Repeat the process until all vertices are ordered.

All the heuristics described above can be implemented in polynomial time in the size of the graph but are not guaranteed to return optimal solutions. In fact the upper bounds on treewidth returned by them can be exponentially worse. Other polynomial time approximation algorithms with constant approximation factors [Amir, 2001, Becker and Geiger, 2001] however perform worse both in terms of cpu time and quality of approximation on real-world problems and random graphs [Amir, 2001]. In our studies and consistent with previous studies [Koster et al., 2001], we found that the min-fill heuristic yields better upper bounds than min-width and max-cardinality heuristic.

## 4 Lower bound on treewidth

---
**Algorithm minor-min-width (G)**
**Input:** A graph G.
**Output:** A lower bound on the treewidth of $G$.

1. lb=0;
2. Repeat
   (a) Contract the edge between a minimum degree vertex $v$ and $u \in N(v)$ such that the degree of $u$ is minimum in $N(v)$ to form a new graph $G'$.
   (b) $lb = MAX(lb, degree_G(v))$.
   (c) Set G to G'.
3. until no vertices remain in G.
4. return lb
---

Figure 1: Algorithm minor-min-width to compute a lower bound on the treewidth of the graph

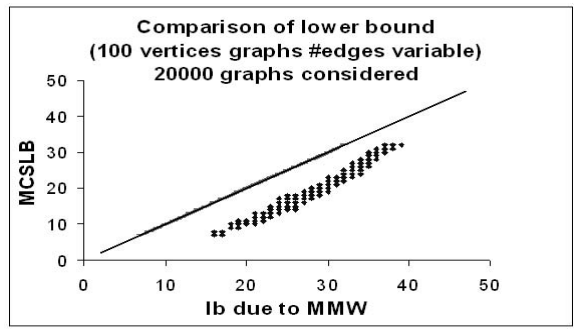

Figure 2: Comparison of Lower bound for graphs with 100 vertices.

In this section, we consider several lower bounds on treewidth. The oldest known lower bound that comes from constraint theory is called the min-width bound [Freuder, 1985] and is based on the min-width heuristic. The idea is that if in a min-width ordering some vertex $v$ has an edge with $lb$ vertices ordered below it in graph $G$, then the treewidth of the graph is at least $lb$. Recently, Brian Lucena [Lucena, 2003] developed a lower bound using the maximum cardinality search and showed that it is almost always better than the min-width bound. The idea is that if in a maximum cardinality ordering, some vertex $v$ has $lb$ vertices which are both ordered below $v$ and adjacent to $v$, then the treewidth of $G$ is at least $lb$. Hence forth, we will refer to this bound as $MCSLB$.

We develop a new lower bound that improves upon the min-width bound by using a celebrated theorem from graph theory which states that the treewidth of a graph is never less than the treewidth of its minor. The resulting algorithm which we call minor-min-width (MMW) is given in Figure 4. We observe em-

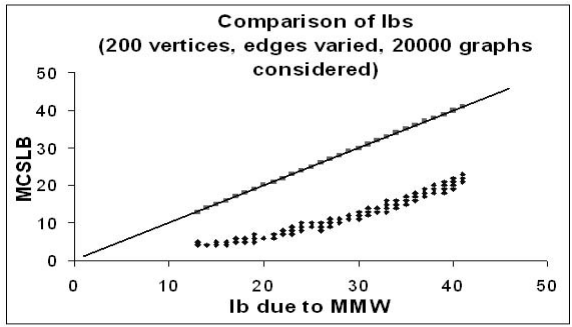

Figure 3: Comparison of Lower bound for graphs with 200 vertices.

---

**Algorithm Treewidth Branch and Bound(G)**
**Input:** A graph $G$.
**Output:** Treewidth of $G$.

1. **Initialize:** A state $s$ which is a two-tuple consisting of a graph $G^s = G$ and a partial order $x^s = \phi$, $g(s) = 0$, $h(s) = mmw(G)$, $f(s) = h(s)$. Upper bound $ub$ is set to the bound computed by the min-fill algorithm.

2. If$(f(s) < ub)$ **BB**$(s)$

3. return $ub$

**sub-procedure BB(s)**

1. IF $|V_{G^s}| < 2$ THEN $ub = MIN(ub, f(s))$

2. ELSE FOR each vertex $v$ in $G^s$ do
    (a) Create a state $s' = (G^{s'}, x^{s'})$ where $G^{s'} = elim(G^s, v)$ and $x^{s'} = (x^s, v)$.
    (b) $g(s') = MAX(g(s), degree_{G^s}(v))$
    (c) $h(s') =$ **minor-min-width**$(G^{s'})$
    (d) $f(s') = MAX(g(s'), h(s'))$
    (e) If $f(s') < ub$ then $BB(s')$

---

Figure 4: Algorithm Treewidth Branch and Bound

pirically that the bound computed by MMW is almost always better than MCSLB. This is shown in Figures 2 and 3. Figure 2 is a scatter plot of MCSLB and MMW for 100 vertex random graphs while Figure 3 is a scatter plot for 200 vertex random graphs. These plots show a total domination of the $MMW$ bound over $MCSLB$. We can show that:

**Theorem 4.1.** *Minor-min-width computes a lower bound on the treewidth of the graph.*

## 5 The Branch and Bound Algorithm

The naive branch and bound algorithm given in Figure 4 operates as follows. First it computes the upper bound on treewidth of the graph by using the min-fill heuristic. Next, it computes a lower bound by using the MMW algorithm described above. If the lower bound equals the treewidth returned by the min-fill algorithm, it is returned as the optimal solution. Otherwise, we initialize the best solution found so far to the min-fill solution and start a branch and bound search for a better solution. Once a partial solution is found whose lower bound on treewidth is greater than $ub$, we prune the branch of the search. On the other hand, if we find a complete ordering that is better than the best so far, we update the best solution found so far (stored in $ub$). Note that the algorithm performs a search in the space of perfect elimination ordering of vertices of $G$.

Each state $s$ in the algorithm is a two-tuple: a graph $(G^s)$ and a partial order $x^s$. The graph $G^s$ at state $s$ is obtained by eliminating the vertices along the partial order $x^s$ from the original graph $G$. The successors of a state $s$ can be obtained by eliminating a vertex $v \in V_{G^s}$ of the graph $G^s$. The $g$ value of a state $s$ is the width of the ordering $x^s$ along the path from the root while its $h$ value is the lower bound on the treewidth of $G^s$. We say that a state $s'$ eliminates a vertex $v$ if it is created from state $s$ according to step $3(a)$ in Figure 4.

**Theorem 5.1.** *When algorithm* **Treewidth Branch and Bound** *terminates, $ub$ stores the treewidth of $G$.*

We improve upon this Branch and Bound algorithm in three ways: (1) Improving the $f$ value at each state (2) Reducing the branching factor at each state (3) Using propagation and pruning rules (discussed in next section).

### 5.1 Graph reduction techniques

Graph reduction techniques are based on the intuition that we can delete some vertices from the graph without affecting its treewidth. In our branch and bound setting, this translates to adding some vertices to the partial order $x^s$ and reducing the size of the graph $G^s$ at a state $s$. We use two graph reduction techniques called *the simplicial vertex rule* and the *almost simplicial vertex rule* due to Bodlaender et al. [Bodlaender et al., 2001]. The simplicial vertex rule states that if vertex $v$ is simplicial in graph $G$, then treewidth of $G$ is $MAX(degree_G(v), tw(elim(G, v)))$. Similarly, the almost simplicial vertex rule states that if $v$ is a almost simplicial vertex in graph $G$ and $degree_G(v) \leq lb$, then treewidth of $G$ is $tw(elim(G, v))$ where $lb$ is a lower bound on the treewidth of $G$. Hence forth, we will abuse notation and use the term almost simplicial vertices to mean those vertices that are almost simplicial and have degree less than the lower bound $lb$. We can incorporate these rules in the algorithm given in Figure 4 using the following pseudocode (after step 3(d) in sub-procedure BB(s)). Let $v$ be a simplicial or almost simplicial vertex in graph $G^s$ at state $s$.

Repeat
1. Update $s$: $G^s = elim(G^s, v)$, $x^s = (x^s, v)$
2. $g(s) = Max(g(s), degree_{G^s}(v))$.
3. $f(s) = MAX(g(s), f(s))$

until $G^s$ has no simplicial or almost simplicial vertices.

By applying graph reduction techniques, we achieve two things. First, we may reduce the branching factor at each state because we are now branching on all vertices of a potentially smaller graph. Secondly, the value of $g$ at state $s$ may increase providing us with more pruning opportunities. Note that one could easily prove that the resulting algorithm will correctly compute the treewidth of the graph by using results from [Bodlaender et al., 2001].

### 5.2 Simplicial vertices and chordal graphs

In this subsection, we show that at any state $s$, one needs to consider only the non-neighbors of the currently ordered vertex $v$ as successors of $s$.

**Definition 5.2.** *Let $P$ be the set of all possible orderings $\pi = (V_1, V_2, \ldots, V_n)$ of vertices of $G$ constructed in the following manner. Select an arbitrary vertex and place it at position 1. For $i = 2$ to $n$, If there exists a vertex $v$ such that $v \notin N(V_{i-1})$, make it simplicial and remove it from $G$. Otherwise, select an arbitrary vertex $v$ and remove it from $G$. Place $v$ at position $i$. $P$ is called the **treewidth elimination set** of $G$.*

**Lemma 5.3.** *Let $P$ be a treewidth elimination set of a graph $G$ and let $tw$ be the treewidth of $G$. There exists an ordering $\pi$ in $P$ such that triangulating $G$ along the ordering $\pi$ will result in a triangulation of treewidth $tw$.*

Lemma 5.3 can be proved using graph minor theory and Dirac's theorem on triangulated graphs (Lemma 2.7). Based on this lemma, we can replace the *for* statement in sub-procedure $BB(s)$ in Figure 4 by "For $v \notin N(v^s)$ do" where $v^s$ is the last vertex in partial ordering $x^s$.

### 5.3 The Edge addition rule

In this subsection, we show that we can add a set of edges to the graph $G^s$ at any state $s$ without sacrificing correctness. We achieve this using the following theorem.

**Theorem 5.4.** *Let $G(V, E)$ be a graph. If $ub$ is a upper bound on the treewidth of $G$ and there exists two vertices $v_1$ and $v_2$ in $G$ such that $|N(v_1) \cap N(v_2)| \geq ub + 1$, then there must be an edge between $v_1$ and $v_2$ in all possible perfect elimination orderings of $G$ that have treewidth less than or equal to $ub$.*

Informally, Theorem 5.4 can be incorporated into the algorithm treewidth branch and bound as follows. At each state $s$, if we find two vertices that have more than $ub + 1$ common neighbors then we connect them in the graph $G^s$. Again by adding new edges, we can potentially increase the pruning power and reduce the branching factor by creating new simplicial and almost simplicial vertices.

## 6 Propagation and Pruning Rules

In this section, we will state a set of theorems which will help us prune regions of the search space that would not result in triangulations of better treewidth than the current best. We will skip the proofs due to space constraints. Before we proceed, let us first introduce some terminology commonly used in AI literature. A state $s$ is said to be explored when all its children are visited. The descendants of a state $s$ can be defined recursively as follows: (1) All children of a state are descendants of $s$. (2) If $c$ is descendant of $b$ and $b$ is descendant of $a$, then $c$ is a descendant of $a$. We say that the algorithm treewidth branch and bound is correct iff it outputs the treewidth of the graph provided to it as input.

**Theorem 6.1.** *Let $A$ and $B$ be two vertices in graph $G^s$ at state $s$. Let $s_a$ be the child of $s$ that eliminates $A$ and $s_{ab}$ be the child of $s_a$ that eliminates $B$. Also, Let $s_b$ be the child of $s$ that eliminates $B$ and $s_{ba}$ be the child of $s_b$ that eliminates $A$. If Treewidth branch and bound explores $s_{ab}$ and prunes $s_{ba}$, then it is correct.*

**Theorem 6.2.** *Let $A$ be a vertex in graph $G^s$ at state $s$. Let $s_a$ be the child of $s$ that eliminates $A$. Let $s'$ be a descendant of $s$ that eliminates $A$ such that $N_{G^{s_a}}(A) = N_{G^{s'}}$. If Treewidth branch and bound explores state $s_a$ and prunes $s'$, then it is correct.*

**Theorem 6.3.** *Let $A$ and $B$ be two vertices in graph $G^s$ at state $s$ such that eliminating vertex $A$ makes vertex $B$ simplicial or almost simplicial and eliminating vertex $B$ makes vertex $A$ simplicial or almost simplicial. Let $s_a$ be the child of $s$ that eliminates $A$ and let $s_b$ be the child of $s$ that eliminates $B$. If Treewidth branch and bound explores $s_a$ and prunes $s_b$, then it is correct.*

**Theorem 6.4.** *Let $A$ and $B$ be two vertices in graph $G^s$ at state $s$ such that $E(A) \subseteq E(B)$ where $E_A$ and $E_B$ are the set of edges added respectively when the operations $elim(G^s, A)$ and $elim(G^s, B)$ are carried out on graph $G^s$. Let $s_a$ be the child of $s$ that eliminates $A$ and let $s_b$ be the child of $s$ that eliminates $B$. If Treewidth branch and bound explores $s_a$ and prunes $s_b$, then it is correct.*

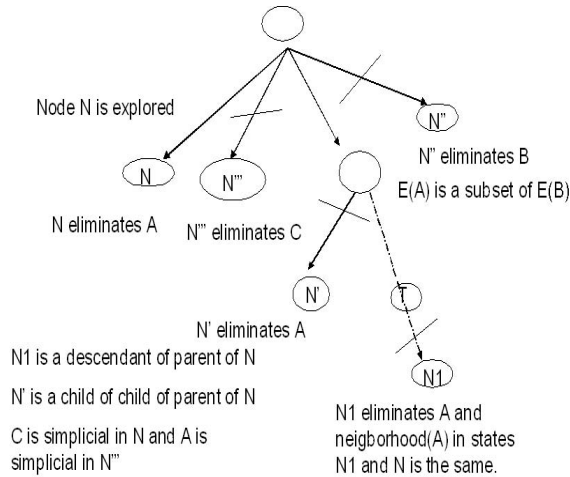

Figure 5: States $N'$, $N1$, $N'''$ and $N''$ are pruned by Theorems 6.1, 6.2, 6.3 and 6.4 respectively. Pruned nodes are indicated by drawing a "-" through them.

All the theorems stated in this section can be easily incorporated into the algorithm described in Figure 4. A pictorial view of the pruning caused by these theorems is given in Figure 5. We call the algorithm resulting from the addition of heuristics and pruning methods discussed in this and the previous section as *QuickBB*.

## 7 Experimental Results

In this section, we present experimental results on running QuickBB on random graphs, randomly generated partial k-trees and benchmarks from Bayesian networks repository and Second Dimacs graph coloring challenge. For comparison, we also solve each problem by min-fill heuristic and Shoikhet et al.'s implementation [Shoikhet and Geiger, 1997] of the QuickTree algorithm. The branch and bound algorithm was implemented using C++ and STL. All experiments were run on a Pentium-4 2.4 GHz machine having 2 GB of RAM. We have implemented a randomized version of the min-fill algorithm and in all the experiments reported below, we consider only the lowest treewidth computed in 100 runs of this randomized algorithm. Shoikhet et al.'s implementation (QuickTree) which solves the decision version of the treewidth problem requires a value of $k$ to be provided as input. So, we first compute an upper bound $ub$ (or optimal if known) on the treewidth of the graph using QuickBB and provide this as an input to QuickTree. Thus, we are running QuickTree with favorable settings.

The tables use the following terminology. The columns for QuickBB and QuickTree are labeled as QBB and QT respectively. The column Tw gives the treewidth output by QuickBB. The column LB gives the lower bound output by the minor-min-width algorithm. The column MF gives the treewidth computed by the min-fill heuristic after 100 iterations. The column Nodes gives the number of nodes explored by the QuickTree algorithm. Finally, a "*" indicates that the algorithm did not terminate.

Table 1: Random Graphs

| N | E | Time | | Nodes | Tw | LB | MF |
|---|---|---|---|---|---|---|---|
| | | QBB | QT | | | | |
| 25 | 50 | 5.3671 | 55.3113 | 6406 | 6.3 | 5.5 | 6.4 |
| 25 | 100 | 33.14 | 68.6511 | 12398 | 11.8 | 9.5 | 12 |
| 25 | 150 | 24.6956 | 33.2384 | 4228 | 15.1 | 11.7 | 15.5 |
| 25 | 200 | 3.2196 | 13.1456 | 493 | 18 | 17.2 | 18.5 |
| 50 | 50 | 0.654 | 27.43 | 76.5 | 3.6 | 3.1 | 3.6 |
| 50 | 100 | 201.34 | * | 56898 | 10.8 | 8.2 | 11.2 |
| 50 | 150 | 745.3 | * | 134553 | 17.3 | 11.8 | 17.6 |
| 50 | 200 | 1856.7 | * | 345678 | 20.34 | 12.3 | 22.3 |
| 50 | 300 | 3267.21 | * | 445789 | 27.6 | 13.8 | 29.3 |
| 50 | 400 | 2674.3 | * | 345678 | 33.7 | 16.2 | 35.2 |
| 50 | 500 | 855.23 | * | 91232 | 34.7 | 19.35 | 34.9 |
| 50 | 600 | 876.5 | * | 85621 | 37.51 | 22.3 | 37.8 |
| 50 | 700 | 499.4 | 2345.67 | 50929 | 39.5 | 25.8 | 39.6 |
| 50 | 800 | 203.4 | 229.87 | 23445 | 41.2 | 28.7 | 41.2 |
| 50 | 900 | 33.45 | 121.23 | 8922 | 42.3 | 32.5 | 42.4 |
| 50 | 1000 | 5.43 | 36.7 | 453 | 44.2 | 35.6 | 44.7 |

### 7.1 Random graphs

All the graphs used in this subsection were generated using the parametric model $(n, m)$, where $n$ is the number of vertices in the graph and $m$ is the number of edges. Here, we select $m$ edges uniformly at random from the possible set of $n * (n - 1)/2$ edges to create a random graph $G$. Table 1 shows the results of running min-fill, QuickTree and QuickBB on randomly generated graphs having 25 and 50 vertices respectively. Each algorithm was given a maximum of 1 and 4 hours respectively on 25 and 50 vertex graphs after which a time-out was reported. The values in the table are averages over 100 instances for each combination of $m$ and $n$.

We can see that QuickBB is almost always faster than QuickTree. Also, note that the min-fill algorithm always yields close to optimal treewidth for these graphs. The largest difference between the average treewidth output by min-fill and the average optimal value was never more than 2 and 0.5 for 50-vertex and 25-vertex random graphs respectively. We observed that graphs of small (close to 1) and large (close to $n$) treewidth are easy for both QuickTree and QuickBB while graphs of intermediate treewidths are harder. Note that QuickBB solved all 50-vertex graphs in less than 2 hours each.

### 7.2 Randomly generated partial k-trees

All the graphs used in this subsection were generated using the parametric model $(n, k, p)$ that generates partial $k$-trees as follows. We first generate a random $k$-tree having $n$ vertices by first forming a clique of size $k + 1$. We then add the remaining $n - k - 1$ vertices by making the new vertex adjacent to a clique of size $k$ selected uniformly at random from the cliques already present in the graph. Then, we remove $p$ percent edges from this $k$-tree uniformly at random to form a partial $k$-tree. Table 2 shows the results of running various algorithms on randomly generated partial $k$-trees having 50, 100 and 200 vertices. $p$ was varied from 20 to 60 in increments of 20. $k$ was set to 10. The time-out for each algorithm was set to 4 hours. The values reported in the table are averages on 100 problem instances for each combination of $n$, $k$ and $p$ considered. Again, we observe that QuickBB is almost always faster than QuickTree.

Table 2: Random Partial K-trees

| N | K | P | Tw | Time QBB | Time QT | Nodes | LB | MF |
|---|---|---|---|---|---|---|---|---|
| 50 | 10 | 20 | 10.0 | 0.4 | 6.1 | 94.4 | 9.2 | 10.3 |
| 50 | 10 | 40 | 10.0 | 0.6 | 32.7 | 157.4 | 9.5 | 10.3 |
| 50 | 10 | 60 | 9.9 | 29.6 | 103.1 | 1999.2 | 9.1 | 10.0 |
| 100 | 10 | 20 | 10.0 | 1.5 | 259.7 | 206.3 | 10.0 | 10.5 |
| 100 | 10 | 40 | 10.0 | 1.7 | * | 238.7 | 9.7 | 10.8 |
| 100 | 10 | 60 | 9.6 | 103.2 | * | 5167.5 | 8.9 | 10.3 |
| 200 | 10 | 20 | 10.0 | 15.6 | * | 791.1 | 10.0 | 10.7 |
| 200 | 10 | 40 | 10.0 | 8.8 | * | 728.3 | 9.6 | 10.4 |
| 200 | 10 | 60 | 9.9 | 3.2 | * | 261.7 | 9.2 | 11.0 |

The purpose of this study was to evaluate the performance of QuickBB on graphs having bounded treewidth. It can be seen that the practical limit for the QuickTree implementation is the partial $k$-trees generated by the parametric model $(100, 10, 20)$. On the other hand, QUICKBB solved all partial $k$-trees generated by the parametric model $(200, 10, p)$ for $p = 20, 40, 60$ in less than one hour of cpu time.

Table 3: Networks In Bayesian Network Repository

| Network | V | E | Tw | Time | Nodes | LB | MF |
|---|---|---|---|---|---|---|---|
| alarm | 38 | 65 | 4 | 0.017054 | 1 | 4 | 4 |
| barley | 49 | 126 | 7 | 48.7506 | 14588 | 6 | 7 |
| diabetes | 414 | 819 | 4 | 206.023 | 9385 | 4 | 4 |
| **link** | **715** | **1738** | **13** | * | **136190** | **8** | **15** |
| mildew | 36 | 80 | 4 | 0.113972 | 35 | 4 | 4 |
| munin1 | 190 | 366 | 11 | * | 373690 | 8 | 11 |
| munin2 | 1004 | 1662 | 7 | * | 40421 | 6 | 7 |
| munin3 | 1045 | 1745 | 7 | 109.284 | 2000 | 7 | 7 |
| munin4 | 1042 | 1843 | 8 | * | 61138 | 7 | 8 |
| oescoa42 | 43 | 72 | 3 | 0.019563 | 1 | 3 | 3 |
| oesoca | 40 | 67 | 3 | 0.018448 | 1 | 3 | 3 |
| oesoca+ | 68 | 208 | 11 | 2.78736 | 491 | 9 | 11 |
| pathfinder | 110 | 211 | 6 | 0.201469 | 25 | 6 | 6 |
| pigs | 442 | 806 | 10 | * | 182365 | 7 | 10 |

### 7.3 Bayesian networks

We also generated statistics on some well known graphs in the Bayesian network repository[1]. We were able to compute the treewidth of some graphs whose optimal treewidth was not yet known. The results are shown in Table 3. The time-bound used was 1 hour.

[1] www.cs.huji.ac.il/labs/compbio/Repository/

We were able to compute the optimal treewidth for the following networks: Alarm, Barley, Diabetes, mildew, munin2, oescoa42, oesoca and pathfinder. While on other networks like Link, munin (1,3 and 4) and pigs the algorithm ran out of time. It is interesting to note that we were able to improve upon the treewidth output by the min-fill algorithm only in the case of the Link network. For other networks on which QuickBB terminates, the treewidth output by the min-fill algorithm was equal to the optimal value.

Table 4: Dimacs Graph Coloring Instances

| Graph | V | E | Tw | Tw* | Time | LB |
|---|---|---|---|---|---|---|
| anna | 139 | 986 | 12 | 12 | 1.64 | 11 |
| david | 88 | 812 | 13 | 13 | 77.6538 | 11 |
| huck | 75 | 602 | 10 | 10 | 0.041 | 10 |
| homer | 557 | 3258 | 31 | 31 | * | 19 |
| jean | 78 | 508 | 9 | 9 | 0.05 | 9 |
| queen5-5 | 26 | 320 | 18 | 18 | 5.409 | 12 |
| **queen6-6** | **37** | **580** | **25** | **26** | **81.32** | **15** |
| queen7-7 | 50 | 952 | 35 | 35 | 543.3 | 18 |
| queen8-8 | 65 | 1456 | 46 | 46 | * | 22 |
| queen9-9 | 82 | 2112 | 59 | 59 | * | 25 |
| **queen10-10** | **101** | **2940** | **72** | **73** | * | **29** |
| queen11-11 | 122 | 3960 | 89 | 89 | * | 34 |
| queen12-12 | 145 | 5192 | 110 | 106 | * | 38 |
| queen13-13 | 170 | 6656 | 125 | 125 | * | 42 |
| **queen14-14** | **197** | **8372** | **143** | **145** | * | **47** |
| queen15-15 | 226 | 10360 | 167 | 167 | * | 51 |
| queen16-16 | 257 | 12640 | 205 | 191 | * | 56 |
| fpsol2.i.1 | 270 | 11654 | 66 | 66 | 0.587076 | 66 |
| fpsol2.i.2 | 364 | 8691 | 31 | 31 | 0.510367 | 31 |
| fpsol2.i.3 | 364 | 8688 | 31 | 31 | 0.492061 | 31 |
| inithx.i.1 | 520 | 18707 | 56 | 56 | 26.3043 | 55 |
| **inithx.i.2** | **559** | **13979** | **31** | **35** | **1.05661** | **31** |
| **inithx.i.3** | **560** | **13969** | **31** | **35** | **1.02734** | **31** |
| miles1000 | 129 | 6432 | 49 | 49 | * | 45 |
| miles1500 | 129 | 10396 | 77 | 77 | 6.759 | 77 |
| **miles250** | **126** | **774** | **9** | **10** | **1.788** | **9** |
| miles500 | 129 | 2340 | 22 | 22 | 1704.62 | 21 |
| miles750 | 129 | 4226 | 37 | 37 | * | 33 |
| mulsol.i.1 | 139 | 3925 | 50 | 50 | 1.407 | 50 |
| mulsol.i.2 | 174 | 3885 | 32 | 32 | 3.583 | 32 |
| mulsol.i.3 | 175 | 3916 | 32 | 32 | 3.541 | 32 |
| mulsol.i.4 | 176 | 3946 | 32 | 32 | 3.622 | 32 |
| mulsol.i.5 | 177 | 3973 | 31 | 31 | 3.651 | 31 |
| myciel3 | 12 | 20 | 5 | 5 | 0.059279 | 4 |
| **myciel4** | **24** | **71** | **10** | **11** | **0.205416** | **8** |
| **myciel5** | **48** | **236** | **19** | **20** | **112.12** | **14** |
| myciel6 | 96 | 755 | 35 | 35 | * | 23 |
| **myciel7** | **192** | **2360** | **54** | **69** | * | **39** |
| **le450-5a** | **451** | **5714** | **307** | **308** | * | **53** |
| **le450-5b** | **451** | **5734** | **309** | **313** | * | **52** |
| **le450-5c** | **451** | **9803** | **315** | **340** | * | **75** |
| **le450-5d** | **451** | **9757** | **303** | **326** | * | **73** |
| **le450-15b** | **451** | **8169** | **289** | **296** | * | **59** |
| **le450-15c** | **451** | **16680** | **372** | **376** | * | **98** |
| le450-15d | 451 | 16750 | 371 | 371 | * | 96 |
| le450-25a | 451 | 8260 | 255 | 255 | * | 57 |
| le450-25b | 451 | 8263 | 251 | 251 | * | 54 |
| **le450-25c** | **451** | **17343** | **349** | **355** | * | **97** |
| **le450-25d** | **451** | **17425** | **349** | **356** | * | **95** |
| DSJC1000.1 | 1001 | 99258 | 896 | * | * | 183 |
| DSJC1000.5 | 1001 | 499652 | 977 | * | * | 469 |
| DSJC1000.9 | 1001 | 898898 | 991 | * | * | 872 |
| **DSJC125.1** | **126** | **1472** | **64** | **67** | * | **20** |
| **DSJC125.5** | **126** | **7782** | **109** | **110** | * | **56** |
| DSJC125.9 | 126 | 13922 | 119 | 119 | 260.879 | 104 |
| **DSJC250.1** | **251** | **6436** | **176** | **179** | * | **43** |
| **DSJC250.5** | **251** | **31336** | **231** | **233** | * | **114** |
| DSJC250.9 | 251 | 55794 | 243 | 243 | * | 212 |
| DSJC500.1 | 501 | 24916 | 409 | * | * | 87 |
| DSJC500.5 | 501 | 125248 | 479 | * | * | 231 |
| DSJC500.9 | 501 | 224874 | 492 | * | * | 433 |
| DSJR500.1c | 501 | 242550 | 485 | * | 656.198 | 474 |
| DSJR500.5 | 501 | 117724 | 175 | * | * | 176 |

### 7.4 Dimacs Graph coloring networks

The purpose of these experiments was to test the anytime properties of our algorithm on large instances. The results are shown in Table 4. The column Tw* in Table 4 is the best upper bound on treewidth reported by Koster et al. [Koster et al., 2001] in their rigorous computational study on heuristics like min-fill, max-cardinality search and a local-search like procedure called minimum-separating-vertex-set (MSVS) heuristic. QuickBB was given a maximum of 3 hours of cputime on each instance. We observe that QuickBB was able to improve upon previously known upper bounds on most instances. On some instances like myceil7 and le450-5c the improvement was dramatic($\approx 20$) while on other instances like DSJC250.5 and DSJC125.1 the improvement was minor($\approx 2$). We also ran QuickTree algorithm (not reported in the table) on these instances and found that QuickTree did not terminate except on the following four instances: myceil3, myceil4, queen5-5 and queen6-6.

## 8 Summary and Future work

We developed a complete anytime algorithm to compute the treewidth of a graph. In the course of the development of this algorithm, we were able to improve upon the recently developed lower bound on the treewidth of a graph by Brian Lucena [Lucena, 2003]. Experimental results suggest the promise of our approach in that we are consistently able to compute tree-decompositions having smaller treewidth than those computed by the polynomial algorithms like the min-fill heuristic. Also our algorithm scales better than QuickTree(Tables 1 and 2). Another important property of our algorithm is its anytime nature and it is evident from the results on Dimacs graph coloring instances on which a consistent improvement in the upper bound was achieved with time.

However several avenues remain for future work. The algorithm developed by Arnborg et al. [Arnborg et al., 1987] has a time complexity of $O(n^{k+2})$ where $n$ is the number of vertices of the graph and $k$ is the treewidth of the graph. The worst case time complexity of QuickBB scales as $O(n^{n-k})$. In other words, QuickBB favors problems having higher treewidth and is evident from Table 1. We believe that QuickBB could be improved so that its time complexity is $O(min(n^k, n^{n-k}))$. Other avenues for future work include conducting rigorous empirical tests on benchmarks like csplib, satlib and random probabilistic networks.

### Acknowledgments

This work was supported in part by the NSF grant IIS-0086529 and the MURI ONR award N00014-00-1-0617.